\documentclass[aps,prd,twocolumn,groupedaddress,nofootinbib,amssymb]{revtex4}
\usepackage[latin1]{inputenc}
\usepackage{graphicx}
\usepackage[english]{babel}
\usepackage{amsmath}
\usepackage{amssymb}
\usepackage{amsfonts}

\begin{document}
\def\pp{{\, \mid \hskip -1.5mm =}}
\def\cL{{\cal L}}
\def\be{\begin{equation}}
\def\ee{\end{equation}}
\def\bea{\begin{eqnarray}}
\def\eea{\end{eqnarray}}
\def\beq{\begin{eqnarray}}
\def\eeq{\end{eqnarray}}
\def\tr{{\rm tr}\, }
\def\nn{\nonumber \\}
\def\e{{\rm e}}

\title{\textbf{Coalescing binaries as possible standard candles}}

\author{Salvatore Capozziello, Mariafelicia  De Laurentis, Ivan De Martino, Michelangelo Formisano}

\affiliation{\it Dipartimento di Scienze Fisiche, Università
di Napoli {}``Federico II'', INFN Sez. di Napoli, Compl. Univ. di
Monte S. Angelo, Edificio G, Via Cinthia, I-80126, Napoli, Italy}

\date{\today}

\begin{abstract}
Gravitational waves detected  from well-localized inspiraling
binaries would allow  to  determine, directly and independently,
both   binary  luminosity  and  redshift. In this case, such
systems could behave as "standard candles"  providing  an
excellent probe of cosmic distances up to $z <0.1$ and thus
complementing other indicators of cosmological distance ladder.
 \end{abstract}
 \maketitle {\it Keywords}: gravitational waves, standard
candles, cosmological distances

\section{Introduction}

A new type of standard candles, or, more appropriately, standard
sirens, could be achieved by studying coalescing binary systems
\cite{sirene, sathya}. These systems are usually considered strong
emitter of gravitational waves (GW), ripples of space-time due to
the presence of accelerated masses  in analogy with the
electromagnetic waves, due to accelerated charged. The coalescence
of astrophysical systems containing  relativistic  objects as
neutron stars (NS), white dwarves (WD) and black holes (BH)
constitute very standard GW sources which could be extremely
useful for cosmological distance ladder if physical features of GW
emission are well determined. These binaries systems, as the
famous PSR 1913+16 \cite{hulse,taylor,weisberg}, have  components
that are gradually inspiralling one over the other   as the result
of energy and angular momentum loss due to (also) gravitational
radiation. As  a consequence the GW frequency is increasing and,
if observed, could constitute a "signature" for the whole system
dynamics. The coalescence of a compact binary system is usually
classified in three stages, which are not very well delimited one
from another, namely the \emph{inspiral phase}, the \emph{merger
phase} and the \emph{ring-down phase}. The merger phase is the
process that proceeds until the collision of the bodies and the
formation of a unique object. Its duration depends on the
characteristics of the originating stars and  emission is
characterized by a frequency damping in the time. In merger phase,
 stars are not modelled as rigid sphere due to the presence of a
convulsive exchange of matter \cite{cutler}. GW emission from
merger phase can only be evaluated using the full Einstein
equations. Because of the extreme strong field nature of this
phase, neither a straightforward application of post-Newtonian
theory nor any perturbation theory is very useful. Recent
numerical work \cite{brugmann,pretorious,baker} has given some
insight into the merger problem, but there are no reliable models
for the waveform of the merger phase up to now. Gravitational
radiation from the ring-down phase is  well known and it can be
described by quasi-normal modes \cite{echevarria}. The relevance
of ring-down phase is described in \cite{cardoso}. Temporal
interval between the inspiral phase and the merger one is called
\emph{coalescing time}, interesting for detectors as the American
LIGO (Laser Interferometer Gravitational-Wave Observatory)
\cite{LIGO} and French/Italian VIRGO \cite{VIRGO}. Coalescence is
a rare event and therefore to see several events per year, LIGO
and/or VIRGO must look far beyond our Galaxy. For example, the
expected rate for a NS-NS system (determined from the observed
population of NS-NS binaries) is found to be $80_{ - 70}^{ + 210}$
Myr$^{-1}$ per galaxy \cite{lorimer}. From this figure, one finds
that the expected rate for the today available sensitivities of
today LIGO and VIRGO is of the order $35_{ - 30}^{ + 90} \times
10^{-3}$ yr$^{-1}$, while for the advanced version of such
interferometers, the rate is more interesting being $190_{ -
150}^{ + 470}$ yr$^{-1}$, that is the probability ranges from  one
event per week to two events per day. A remarkable fact about
binary coalescence is that it can provide an {\it absolute
measurement} of the source distance: this is an extremely
important event in Astronomy.  In fact, for these systems, the
distance is given by the measure of the GW polarization  emitted
during the coalescence. One of the problems which affects the
utilization of coalescing binaries as standard candles is the
measure of the redshift of the source as well as the measure of
the GW polarization  (at present  GWs have not been still
experimentally observed). A solution for the redshift
determination   could be the detection of an electromagnetic
counterpart of the coalescing system (e.g. the detection of an
associated gamma ray burst) or  the redshift measurement of  the
host galaxy or galaxy cluster at their barycenter \cite{koppa08}.
Recent evidence supports the hypothesis that many short-hard
gamma-ray bursts could be  associated with coalescing binary
systems indeed \cite{fox05,nakar06,berger,perley}. In this paper,
we want to show that such systems could be used as reliable
standard candles.

In Sect. \ref{SecII},  we briefly sketch  the GW emission  from
coalescing binary systems in circular orbit at cosmological
distance. In Sect. \ref{SecIV}, we simulate various coalescing
binary systems (WD-WD, NS-NS, BH-BH) at redshift $z < 0.1$ because
of the observational limits of ground-based-interferometers as
LIGO and VIRGO. The goal of the simulation is the measure of the
Hubble constant  and consequently the use of these systems as
standard candles \cite{shutz}. This new type of standard candles
will be able to increase the confidence level on the other
"traditional" standard candles in Astronomy and, moreover, it
could constitute an effective tool to measure distances  at larger
redshifts. Conclusions are drawn in Sect. \ref{SecV}.


\section{Gravitational radiation from a coalescing binary system}
\label{SecII}
 For a detailed exposition of GW theory see, e.g.
\cite{maggiore,buonanno,carroll}. Let us consider here an
isolated, far away and slowly moving source. In this
approximation,  we can write the GW solution as:
\begin{equation}
\bar{h}_{\mu\nu}(t,\textbf{x})= \frac{{2G}}{r}\frac{{d^2 I_{ij} }}{{dr^2 }}(t_R )\, ,
\label{eq:quadr}
\end{equation}
where $I_{ij}$ is the quadrupole momentum tensor of the energy
density of source, conventionally defined as:

\begin{equation}
I_{ij} (t) = \int {y^i y^j } T^{00} (t,\textbf{y})d^3 y\, ,
\end{equation}
a  tensor defined   at any constant time surface, and $t_R$ is the
retarded time \cite{maggiore}. The distance between source and
observer is denoted as $r$.

The gravitational wave produced by an isolated non-relativistic
object is therefore proportional to the second derivative of the
quadrupole momentum of the energy density at the point where the
past light cone of the observer intersects the source. In
contrast, the leading contribution to electromagnetic radiation
comes from the changing dipole momentum of the charge density.

\begin{figure}
\begin{center}
\begin{tabular}{cc}
\includegraphics[width=0.22\textwidth]{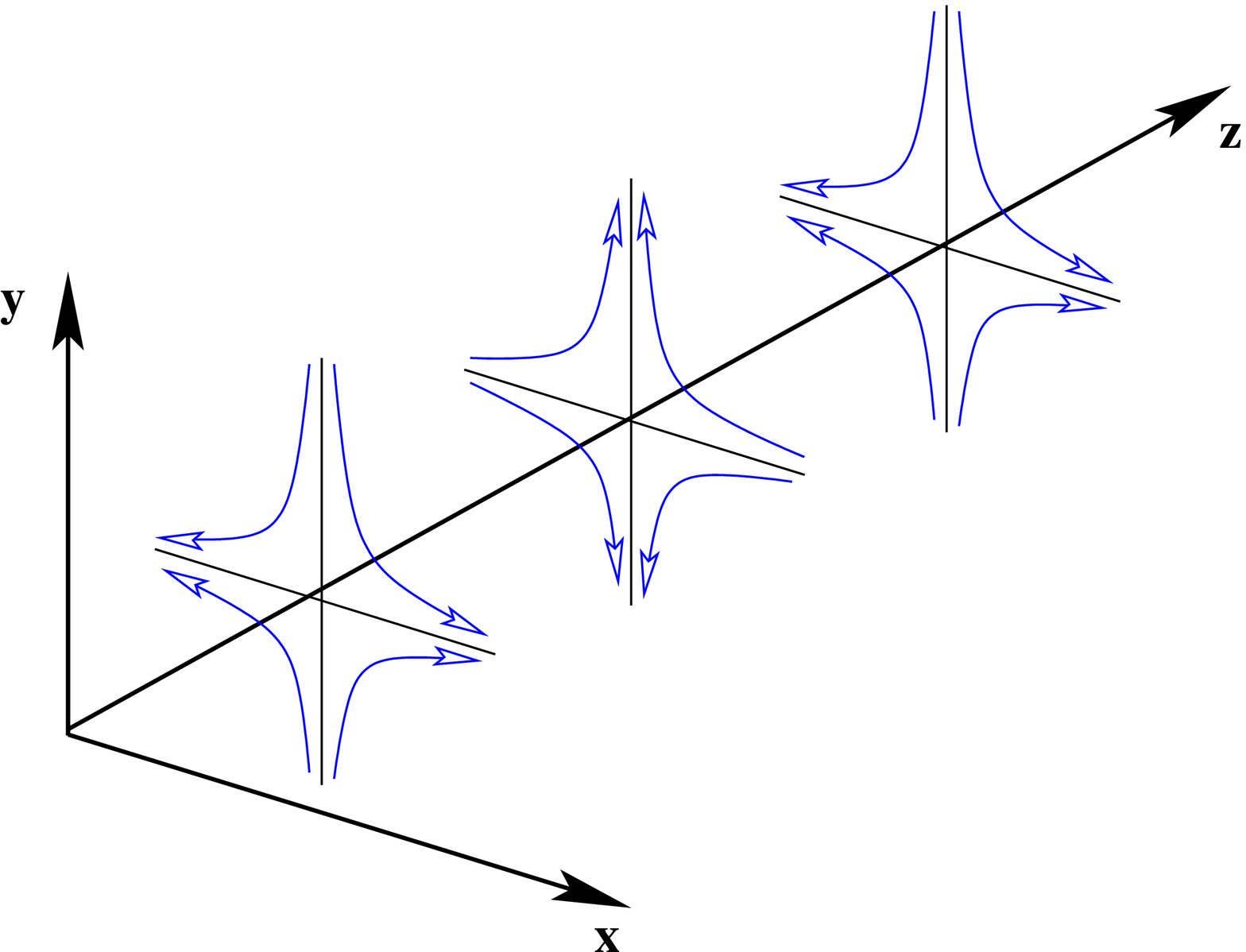} &
\includegraphics[width=0.22\textwidth]{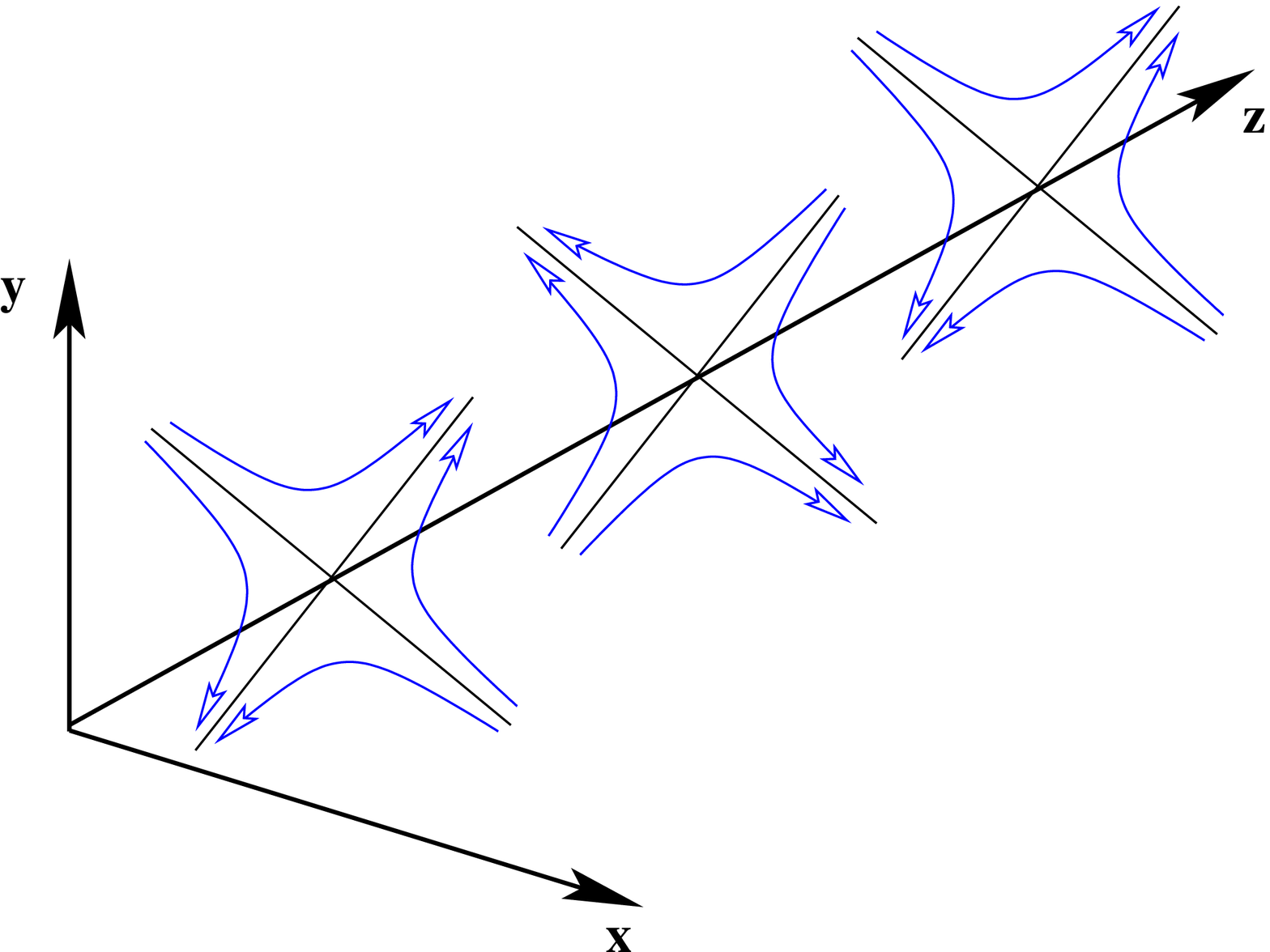} \\
\end{tabular}
\caption{Lines of  force associated to the $+$ (left panel) and
$\times$ (right panel) polarizations.\label{fig:forcelines}}
\end{center}
\end{figure}

The above result can be specified for two point masses $M_1$ and
$M_2$ in a circular orbit. In the quadrupole approximation, the
two polarization amplitudes of GWs at a distance $r$ from the
source are given evaluating Eq.\eqref{eq:quadr} to lowest order in
$v/c$, that is:

\begin{equation}
\begin{array}{l}
 h_ +  \left( t \right) = \dfrac{4}{r}\left( \dfrac{GM_C }{c^2} \right)^{5/3} \left( {\dfrac{{\pi f\left( {t_R } \right)}}{c}} \right)^{2/3} \left( {\dfrac{{1 + \cos ^2 i}}{2}} \right) \times  \\ \\
  \\
 {\rm{            }}\cos \left[ {\Phi \left( {t_R } \right)} \right]\, , \\
 \end{array}
\end{equation}

\begin{equation}
h_ \times  \left( t \right) = \frac{4}{r}\left( {\frac{{G M_C }}{{c^2 }}} \right)^{5/3} \left( {\frac{{\pi f\left( {t_{R} } \right)}}{c}} \right)^{2/3} \cos i\sin \left[ {\Phi \left( {t_{R} } \right)} \right]\\,
\label{eq:polariz}
\end{equation}
where \emph{i} is the binary inclination angle such that $i=90°$
corresponds to a system visible edge-on. These are traditionally
labeled "plus" and "cross" from the lines of force associated
with their tidal stretch and squeeze (see Fig.
\ref{fig:forcelines}).
Here $f$ is  the frequency of the emitted GWs (twice the orbital
frequency). The rate of the frequency change is \cite{buonanno}:

\begin{equation}
\dot f = \frac{{96}}{5}\pi ^{8/3} \left( \frac{GM_C}{c^3 } \right)^{5/3} f^{11/3}\, ,\\
\label{eq:frequenza}
\end{equation}
where \emph{$t_{R}$} is the so called "retarded time" and the
phase \emph{$\Phi$} is given by the expression:
\begin{equation}
\Phi (t) = 2\pi \int\limits_{t_0 }^t {dt'f(t')}\, .\\
\end{equation}
Note that for a fixed distance $r$ and a given frequency $f$, the
GW amplitudes are fully determined by $\mu M^{2/3}  =
{M_C}^{5/3}$, where the combination:

\begin{equation}
M_C  = \frac{{\left( {M_1 M_2 } \right)^{3/5} }}{{\left( {M_1  + M_2 } \right)^{1/5} }}\, ,
\label{eq:Chirp}
\end{equation}
is called \emph{chirp mass} of the binary,  here $M=M_1+M_2$ is
the total mass of the system and $\displaystyle{\mu=\frac{M_1
M_2}{M_1+M_2}}$ is the reduced mass of the system. Introducing the
coalescence time  $\tau  = t_{coal}  - t$ and  integrating  Eq.
\eqref{eq:frequenza}, we get:

\begin{equation}
f \simeq 130\left( \frac{1.21M_ \odot  }{M_C} \right)^{5/8} \left( {\frac{{1\sec }}{\tau }} \right)^{3/8} Hz\, .
\label{eq:evolfreq}
\end{equation}
Eq. \eqref{eq:evolfreq} predicts coalescence times of $\tau\sim17
min,\, 2 sec,\, 1msec$ for $f\sim 10,\,100\,, 1000 Hz$.

After averaging over the orbital period and the orientations of
the binary orbital plane, one arrives at the average
(characteristic) GW amplitude:

\begin{equation}
\begin{array}{l}
 h\left( {f, M_C ,r} \right) = \left( {\left\langle {h_ + ^2 } \right\rangle  + \left\langle {h_ \times ^2 } \right\rangle } \right)^{1/2}  =  \\ \\
 {\rm{                    }} = \left( {\dfrac{{32}}{5}} \right)^{1/2} \dfrac{{G^{5/3} }}{{c^4 }}\dfrac{ M_{C}^{5/3}}{r}\left( {\pi f} \right)^{2/3}\, . \\
 \end{array}
\end{equation}
For a binary at the cosmological distance, i.e. at redshift $z$
where GWs propagate in a Friedmann-Robertson-Walker  Universe,
these equations are modified in a very straightforward way:
\begin{itemize}
\item The frequency that appears in the above formulae is the frequency measured by the observer,
$f_{obs}$, which is red-shifted with respect to the source
frequency $f_s$, i.e. $f_{obs}  = f_s /(1 + z)$, and similarly $t$
and $t_R$ are measured with the observer clocks.
\item The chirp mass $M_C$ has to be replaced by $\mathcal{M}_C  = M_C (1 + z)$.
\item The distance $r$ to the source has to be replaced by the luminosity distance $d_L(z)$.
\end{itemize}
 Inserting the following quantity:

\begin{equation}
h_c \left( t \right) = \frac{4}{{d_L \left( z \right)}}\left( {\frac{{G {\cal{M_C}} \left( z \right)}}{{c^2 }}} \right)^{5/3} \left( {\frac{{\pi f \left( t \right)}}{c}} \right)^{2/3}\, ,
\label{eq:Dis}
\end{equation}
we can rewrite the expressions for the polarization "+" and
"$\times$" as:

\begin{equation}
h_ +  \left( t \right) = h_c \left( {t_{R} } \right)\frac{{1 + \cos ^2 i}}{2}\cos \left[ {\Phi \left( {t_{R} } \right)} \right]\, ,
\end{equation}
and
\begin{equation}
h_ \times  \left( t \right) = h_c \left( {t_{R} } \right)\cos i\sin \left[ {\Phi \left( {t_{R} } \right)} \right]\, .
\end{equation}
Explicating the dependence on  the \emph{chirp mass} redshift, we
can obtain the luminosity distance $d_L$ using the equation
\eqref{eq:Dis} linked directly to the GW polarization:

\begin{equation}
\begin{array}{l}
 d_L \left( z \right) = \dfrac{4}{{h_c \left( t \right)}}\left[ {\dfrac{{GM_C (1 + z)}}{{c^2 }}} \right]^{5/3} \left[ {\pi f(t)} \right]^{2/3}  =  \\
  \\
 {\rm{            }} = \dfrac{{4(1 + z)^{5/3} }}{{h_c \left( t \right)}}\left[ {\dfrac{{G{\cal M}_C }}{{c^2 }}} \right]^{5/3} \left[ {\pi f(t)} \right]^{2/3}\,  . \\
 \end{array}
\label{eq:DistanzaOnde}
\end{equation}
Let us recall that the luminosity distance $d_L$ of a source is defined by
\begin{equation}
{\cal F}=\frac{\cal L}{4\pi d_L^2}\, ,
\end{equation}
where ${\cal F}$ is the  flux (energy per unit time per unit area)
measured by the observer, and  ${\cal L}$ is the absolute
luminosity of the source, i.e. the power that it radiates in its
rest frame. For small redshifts, $d_L$ is related to the present
value of the Hubble parameter $H_0$ and to the deceleration
parameter $q_0$  by
\begin{equation}
\frac{H_0d_L}{c}=z+\frac{1}{2}(1-q_0) z^2+\ldots\, .
\end{equation}
The first term of this expansion gives  the Hubble law $z\simeq
(H_0/c) d_L$, which states that redshift is proportional to the
distance. The term ${\cal O}(z^2)$ is the correction to the linear
law for moderate redshifts. For large redshifts, the Taylor series
is no longer appropriate, and the whole expansion history of the
Universe is encoded in a function $d_L(z)$. For a spatially flat
Universe, one finds
\begin{equation}
d_L(z)=c \, (1+z)\,\int_0^z\, \frac{dz'}{H(z')}\, ,
\end{equation}
where $H(z)$ is the value of the Hubble parameter at redshift $z$.
Knowing $d_L(z)$, we can therefore obtain $H(z)$. This shows that
the luminosity distance function $d_L(z)$ is an extremely
important quantity, which encodes the whole expansion history of
the Universe. Coalescing binaries could be standard candles (or
precisely standard sirens)  in the following sense. Suppose that
we can measure the amplitudes of both polarizations $h_+$,
$h_\times$, as well as $\dot f_{obs}$. From the ratio of $h_+$ and
$h_\times$, we can obtain the value of the inclination of the
orbit, besides, evaluating  $\dot f_{obs}$ at a given frequency,
we can obtain $\mathcal{M}_C$. If we are capable of measuring the
redshift $z$ of the source, we have found a {\it gravitational}
standard candle since we can obtain the luminosity distance from
Eq. \eqref{eq:DistanzaOnde} and then evaluate the Hubble constant
$H_0$. The difference between gravitational standard candles and
the "traditional" standard candles is that the luminosity distance
is directly linked to the GW polarization  and  there is no
theoretical uncertainty on its determination a part  the redshift
evaluation. Various possibilities have been proposed. Among these
there is the possibility to see an optical counterpart. In
fact, it can be  shown that observations of the GWs emitted by
inspiralling  binary compact systems can be a powerful probe at
cosmological scales. In particular, short GRBs appear related to
such systems and  quite promising as potential GW standard sirens
\cite{Dalal}). On the other hand, the redshift of the binary
system can be associated to the barycenter of the host galaxy or
the galaxy cluster as we are going to do here.
\section{Numerical simulation}
\label{SecIV}
We have simulated various coalescing binary systems at redshifts
$z < 0.1$  In this analysis, we do not consider systematic
errors and errors on redshifts to obviate the absence of a
complete catalogue of such systems. The choice of low redshifts is
due to the observational limits of ground-based interferometers
like VIRGO or LIGO. Some improvements
 are achieved, if we take into account  the future generation of these
 interferometers as
 Advanced VIRGO \cite{advirgo}and Advanced LIGO \cite{adligo}. Advanced VIRGO  is a major upgrade, with the
 goal of increasing  the sensitivity by about one order of magnitude with
 respect to VIRGO in the whole detection band. Such a detector,
 with Advanced LIGO, is expected to see many events every year
 (from 10s to 100s  events/year). For example, a NS-NS coalescence
 will be detectable as far as 300 Mpc. In the simulation presented here,
 sources are slightly out of LIGO-VIRGO band but observable, in principle, with  future interferometers.
 Here, we have used the redshifts taken by NASA/IPAC  EXTRAGALACTIC  DATABASE \cite{Abell},
and we have fixed the redshift  using  $z$ at the
barycenter of the host galaxy/cluster), and the binary
\emph{chirp mass} $M_C$,  typically measured, from the Newtonian
part of the signal at upward frequency sweep, to $\sim 0.04\%$ for
a NS/NS binary and $\sim 0.3\%$ for a system containing at least
one BH \cite{Cutler,Curt}. The distance to the binary $d_L$
("luminosity distance" at cosmological distances) can be inferred,
from the observed waveforms, to a precision $\sim 3/\rho \lesssim
30\%$, where $\rho = S/N$ is the amplitude signal-to-noise ratio
in the total LIGO network (which must exceed about 8 in order that
the false alarm rate be less than the threshold for detection). In
this way, we have fixed  the characteristic amplitude of GWs, and
frequencies are tuned in a range compatible  with such a  fixed
amplitude, then the error on distance luminosity is calculated by the error on the chirp mass with standard error propagation.

 The systems considered are NS-NS, BH-BH and WD-WD. For each of them, a particular
frequency range and a characteristic amplitude (beside the chirp
mass) are fixed. We start with the analysis of NS-NS systems ($M_C
= 1.22 M_{\odot}$) with characteristic amplitude fixed to the
value $10^{-22}$. In Table \ref{tab:tabella1}, we report the
redshift, the value of $h_C$ and the frequency range of systems
analyzed.

\begin{figure}[ht]
\includegraphics[scale = 0.3]{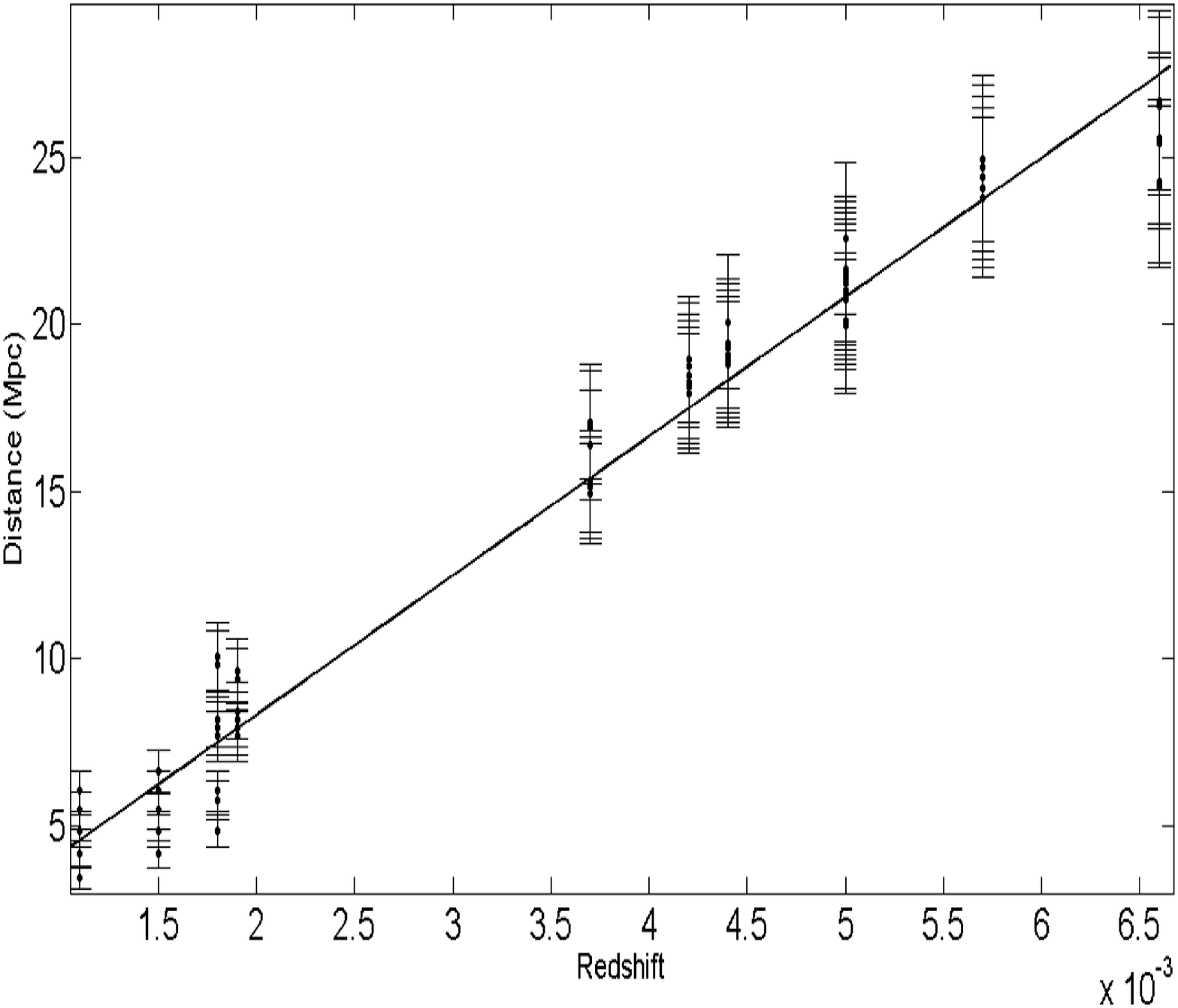}
\caption{Luminosity distance vs redshift for simulated NS-NS systems.}
\label{fig:neutroni}
\end{figure}

\begin{figure}[ht]
\centering
\includegraphics[scale = 0.31]{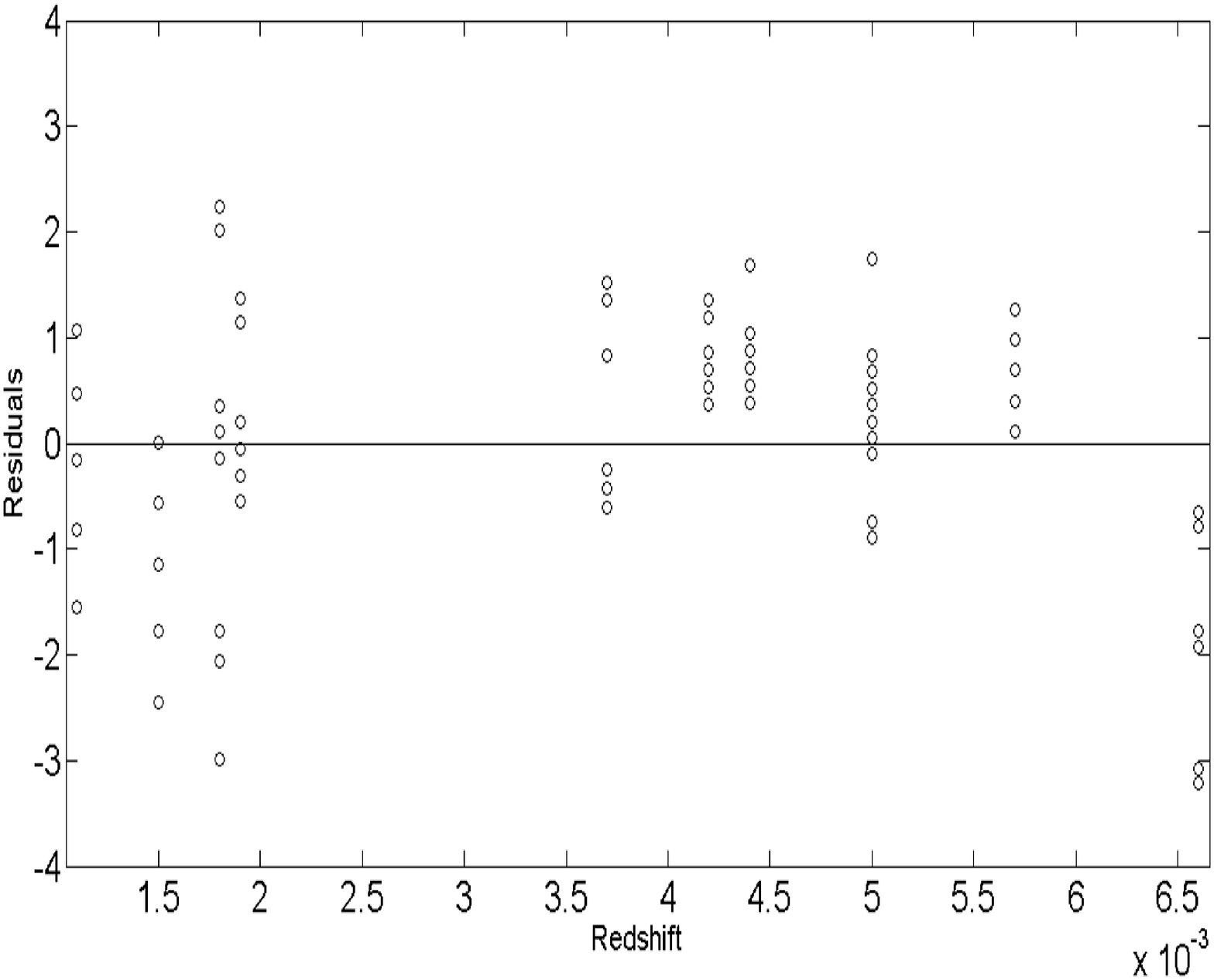}
\caption{Residuals of the plot luminosity distance vs redshift for simulated NS-NS systems.}
\label{fig:ResiduiNeutronStars}

\end{figure}

\begin{table}[htbp]
\centering{
\begin{tabular}{|c|c|c|c|}
  \hline
  \textbf{Object} & \textbf{z} &
  \textbf{$h_c$} & \textbf{Freq. (Hz)} \\
  \hline
  NGC 5128 & 0.0011 & $10^{-22}$ & $0\div10$\\
  \hline
  NGC 1023 Group & 0.0015 & $10^{-22}$ & $0\div10$\\
  \hline
  NGC 2997 & 0.0018 & $10^{-22}$ & $5\div15$\\
  \hline
  NGC 5457 & 0.0019 & $10^{-22}$ & $10\div20$\\
  \hline
  NGC 5033 & 0.0037 & $10^{-22}$ & $25\div35$\\
  \hline
  Virgo Cluster & 0.0042 & $10^{-22}$ & $30\div40$\\
  \hline
  Fornax Cluster & 0.0044 & $10^{-22}$ & $35\div45$\\
  \hline
  NGC 7582 & 0.0050 & $10^{-22}$& $ 45\div55$\\
  \hline
  Ursa Major Groups & 0.0057 & $10^{-22}$& $50\div60$\\
  \hline
  Eridanus Cluster & 0.0066 & $10^{-22}$ & $55\div65$\\
  \hline
    \end{tabular}
\caption {Redshifts, characteristic amplitudes, frequency range for NS-NS systems.}
\label{tab:tabella1}
}
\end{table}
In Fig. \ref{fig:neutroni}, the derived Hubble relation is
reported, and in Fig. \ref{fig:ResiduiNeutronStars}
residuals are reported.

\begin{figure}[h!]
\centering{
\includegraphics[scale=0.3]{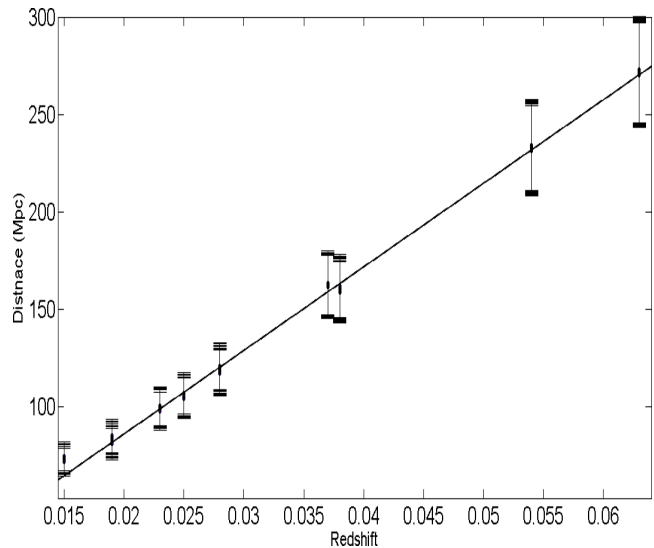}
\caption{Luminosity distance vs redshift for simulated BH-BH systems.}
\label{fig:bh}}
\end{figure}

\begin{figure}[ht]
\centering{
\includegraphics[scale = 0.3]{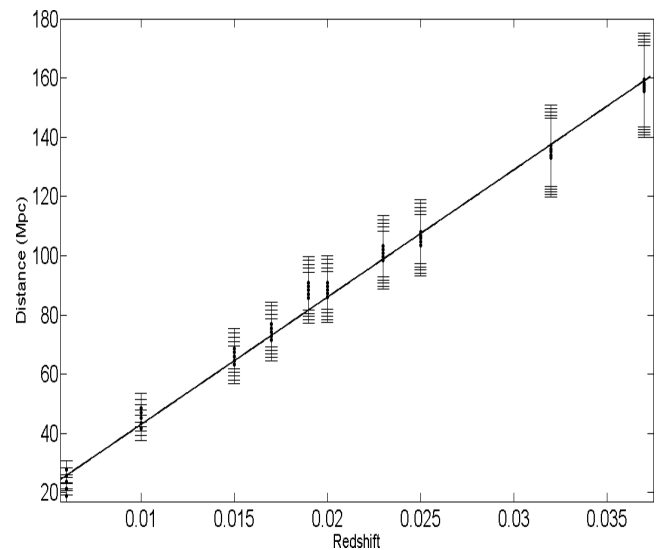}
\caption{Luminosity distance vs redshift for simulated WD-WD systems.}
\label{fig:wd}
}
\end{figure}

The Hubble constant value  is $72\pm1$ $km/sMpc$ in agreement with
the recent WMAP  estimation (Wilkinson Microwave Anisotropy Probe
\cite{WMAP}). The same procedure is adopted for WD-WD systems
($M_C = 0.69 M_{\odot}$, $h_C = 10^{-23}$) and BH-BH systems ($M_C
= 8.67 M_{\odot}$, $h_C = 10^{-21}$). In Tables \ref{tab:tabella2}
and \ref{tab:tabella3}, we report the redshift, the value of $h_C$
and the frequency range for BH-BH and WD-WD systems respectively.
These simulations are reported in Fig. \ref{fig:bh} and in Fig.
\ref{fig:wd}: the corresponding residuals have been checked and their goodness is the same of previous case so we omit them.
\begin{table}
\centering{
\begin{tabular}{|c|c|c|c|}
  \hline
  \textbf{Object} & \textbf{z} &
  \textbf{$h_c$ } & \textbf{Freq. (Hz)} \\
  \hline
  Pavo-Indus Sup.Cluster & 0.015& $10^{-21}$& $ 65\div70$\\
  \hline
  Abell 569 Cluster  & 0.019 & $10^{-21}$ & $ 75\div80$\\
  \hline
  Coma Cluster & 0.023 & $10^{-21}$ & $ 100\div105$\\
  \hline
  Abell 634 Cluster & 0.025 & $10^{-21}$ &$ 110\div115$\\
  \hline
  Ophiuchus Cluster & 0.028 & $10^{-21}$ & $ 130\div135$\\
  \hline
  Columba Cluster & 0.034 & $10^{-21}$ & $ 200\div205$\\
  \hline
  Hercules Sup.Cluster & 0.037& $10^{-21}$& $205\div210$\\
  \hline
  Sculptor Sup.Cluster& 0.054 & $10^{-21}$ & $ 340\div345$\\
  \hline
  Pisces-Cetus Sup.Cluster & 0.063 & $10^{-21}$ & $ 420\div425$\\
  \hline
  Horologium Sup.Cluster & 0.067& $10^{-21}$& $450\div455$ \\
  \hline
    \end{tabular}
\caption {Redshifts, characteristic amplitudes, frequency range for BH-BH systems.}
\label{tab:tabella2}
}
\end{table}

\begin{table}
\centering{
\begin{tabular}{|c|c|c|c|}
  \hline
  \textbf{Object} & \textbf{z} &
  \textbf{$h_c$} & \textbf{Freq. (Hz)} \\
  \hline
  Eridanus Cluster & 0.0066 & $10^{-23}$ &$5\div10$\\
  \hline
  Hydra Cluster & 0.010 & $10^{-23}$ & $15\div20$\\
  \hline
  Payo-Indus Sup.Cluster& 0.015& $10^{-23}$&$35\div40$\\
  \hline
  Perseus-Pisces Sup.Cluster & 0.017 & $10^{-23}$&$40\div45$\\
  \hline
  Abell 569 Cluster & 0.019 & $10^{-23}$ & $45\div50$\\
  \hline
  Centaurus Cluster & 0.020 & $10^{-23}$ & $45\div50$\\
  \hline
  Coma Cluster & 0.023 & $10^{-23}$ & $55\div60$\\
  \hline
  Abell 634 Cluster & 0.025 & $10^{-23}$ & $60\div65$\\
  \hline
  Leo Sup.Cluster & 0.032 & $10^{-23}$ & $85\div90$\\
  \hline
  Hercules Sup.Cluster & 0.037& $10^{-23}$&$100\div105$\\
  \hline
      \end{tabular}
    \caption {Redshifts, characteristic amplitudes, frequency range for WD-WD systems.}
\label{tab:tabella3}
}
\end{table}

For these simulations, the Hubble constant value is $69\pm2$
$km/sMpc$ and $70\pm1$ $km/sMpc$ for BH-BH and WD-WD systems
respectively, also in these cases in agreement with WMAP
estimation.

\section{Conclusions}
\label{SecV}

In this paper, we have considered simulated binary systems
whose redshifts can be estimated considering the barycenter of the
host astrophysical system as  galaxy,  group of galaxies or
cluster of galaxies. In such a way, the standard methods adopted
to evaluate the cosmic distances (e.g. Tully-Fisher or
Faber-Jackson relations) can be considered as "priors" to fit the
Hubble relation. We have simulated various situations assuming
NS-NS, BH-BH, and WD-WD binary systems. Clearly, the leading
parameter is the chirp mass $M_C$, or its red-shifted counter-part
${\cal M}_C$, which is directly related to the GW amplitude. The
adopted redshifts are in a  well-tested  range of scales and  the
Hubble constant value is in good agreement with WMAP estimation.
The Hubble-luminosity-distance diagrams of the above simulations
show the possibility to use the coalescing binary systems as
distance indicators and, possibly, as standard candles.  The
limits of the method are, essentially, the measure of GW
polarizations and redshifts. Besides, in order to improve the
approach, a suitable catalogue of observed coalescing
binary-systems is needed. This is the main difficulty of the
method since, being the coalescence a transient phenomenon, it is
very hard to detect and analyze  the luminosity curves of these
systems. Furthermore, a few simulated sources are out of the
LIGO-VIRGO band.

Next generation of interferometer (as LISA \cite{LISA} or
Advanced-VIRGO and LIGO) could play a decisive role to detect GWs
from these systems. At the advanced level, one expects to detect
at least tens NS-NS coalescing events per year, up to distances of
order $2~Gpc$, measuring the chirp mass with a precision better
than $0.1\%$. The masses of NSs are typically of order
$1.4M_\odot$. Stellar-mass BHs, as observed in X-ray binaries, are
in general more massive, typically with masses of order
$10M_\odot$, and therefore are expected to emit even more powerful
GW signals during their inspiralling and coalescing phases. The
coalescence of two BHs, each one with $10~M_\odot$, could  be seen
by Advanced-VIRGO and Advanced-LIGO up to redshifts $z\sim
2-3$.~\cite{Cutler} Furthermore, the LISA space interferometer,
which is expected to fly in about 10 years, will be sensitive to
GWs in the mHz region, which corresponds to the wave emitted by
supermassive BHs with masses up to $10^6M_\odot$. Nowadays,
supermassive BHs with masses between $10^6$ and $10^9M_\odot$ are
known to exist at the center of most (and probably all) galaxies,
including our Galaxy. The coalescence of two supermassive BHs,
which could take place, for instance during the collision and
merging of two galaxies or in pre-galactic structure at high
redshifts, would be among the most luminous events in the
Universe. Even if the merger rate is poorly understood,
observations from the Hubble Space Telescope and from X-ray
satellites such as Chandra \cite{chandra}  have revealed that
these merging events could be detectable at cosmological
distances. LISA could detect them up to $z\sim
10$,~\cite{Hughes,Vecchio} and it is expected to measure  several
events of this kind. The most important issue that can be
addressed with a measure of $d_L(z)$ is to understand ``dark
energy'', the quite mysterious component of the energy budget of
the Universe that manifests itself through an acceleration  of the
expansion of the Universe at high redshift. This has been
observed, at $z<1.7$, using Type~Ia supernovae as standard
candles~\cite{Riess,Perl}. A possible concern in these
determinations is the absence of a solid theoretical understanding
of the source. After all, supernovae are complicated phenomena. In
particular, one can be concerned about the possibility of an
evolution of the supernovae brightness with redshift, and of
interstellar extinction in the host galaxy  leading to unknown
systematics. GW standard candles  could lead to completely
independent determinations, and complement and increase the
confidence of other standard candles, \cite{HH}, as well as
extending the result to higher redshifts. In the future, the
problem of the redshift could be obviate finding an
electromagnetic counterpart to the coalescence and short GRBs
could play this role.

In summary, this new type of cosmic distance indicators  could be
considered complementary to the traditional standard candles
opening the doors to a self-consistent {\it gravitational
astronomy}.

\acknowledgments
We wish to thank G. Covone and L. Milano for
fruitful discussions and suggestions on the topics of this paper.

\end{document}